\documentclass[twocolumn]{article}
\begin{document}
\rmfamily

\title{\bf
Molecular Bose-Einstein condensate as an amplifier of weak interactions}

\author{V. V. Flambaum and J.S.M. Ginges\\
{\it School of Physics, University of New South Wales,
Sydney 2052,Australia}}
\date{}
\maketitle

\noindent
{\bf {Abstract}}

\noindent
Collisions and chemical reactions of molecules in Bose-Einstein
condensates (BECs) are extremely sensitive to weak fields.
This sensitivity arises due to the high density of compound resonances and
a macroscopic number of molecules with kinetic energy $E=0$ (perfect
energy resolution).
We suggest that chemical reactions in molecular BECs could be used to
enhance effects produced by small external perturbations and search
for a parity-violating energy difference in chiral molecules.

\vspace{0.3cm}

\vspace{0.6cm}

%*************************************************************************
%\thispagestyle{empty}

\noindent
Techniques to trap and cool atoms and molecules are rapidly developing.
Bose-Einstein condensation (BEC) has been realized in
numerous dilute atomic gases \cite{firstBEC}.
While a pure molecular BEC has not yet been achieved,
ultra-cold molecules have been formed \cite{coldmol} and, just
recently, a coherent superposition of an atomic and a molecular
BEC was reported \cite{at-mol}.
\vspace{0.2cm}

\noindent
Collisions of atoms in a BEC can be controlled by the variation of
a relatively weak magnetic field; the magnitude and sign of the
scattering length can be changed by varying the field near a Feshbach
resonance \cite{stwalley,tiesinga}.
In this way the expansion and collapse of an atomic BEC
(termed ``Bosenova'') has been observed \cite{bosenova}.
In collisions of molecules in a BEC, a change of the scattering
length could be achieved with a magnetic field much weaker than
that used to obtain the same effect in atoms.
This is because there is an exponential dependence of the density
of resonances on the number of ``active'' particles.
This high sensitivity to weak fields could be used to search for
a parity violating energy difference between chiral molecules.
\vspace{0.2cm}

\noindent
It is well-known that biological molecules have a definite chiral structure
(for example, there are only naturally occurring left-handed
amino acids and right-handed sugars) \cite{pvreviews}.
There have been numerous attempts to explain this effect
by the influence of the parity violating weak interaction,
which breaks the energy equivalence of right- and left-handed molecules
\cite{rein,letokhov}.
That parity violation can discriminate between molecules of different
chirality is easily seen:
a parity violating nuclear spin-independent electron-nucleus interaction
in atoms \cite{bouchiats,khriplovich} creates a spin helix of the
electrons which interacts differently with right- and left-handed molecules.
However, the parity violating energy difference $\Delta E_{PV}$
is very small \cite{RHS},
\begin{equation}
\label{pved}
\Delta E_{PV}\sim 10^{-20}Z^{5}\eta~{\rm a.u.} \ ,
\end{equation}
where $Z$ is the nuclear charge of the heaviest atom,
and $\eta$ is an asymmetry factor which can be found from
molecular structure calculations. This strong dependence on $Z$ originates
from the weak ($\propto Z^3$) and spin-orbit ($\propto Z^2$) interactions.
It may appear that in molecules with heavy atoms $\Delta E_{PV}$ could
become relatively large due to the $Z^{5}$-dependence.
However, the geometrical suppresion factor $\eta$ remains very small.
We should note that the effect may be larger for molecules with two
heavy atoms \cite{comment2heavy}.
%``1'' and ``2'', where the factor $Z^{5}$ in
%Eq. \ref{pved} should be replaced with the factor $Z_{1}^{3}Z_{2}^{2}$.
For recent calculations of $\Delta E_{PV}$ for various molecules,
see, e.g., \cite{molcalcs} and references therein.
\vspace{0.2cm}

%\noindent
%So far a parity violating energy difference in molecules has eluded
%detection.
%%This is the place to cite experiments searching for an energy difference
%It is difficult to imagine a situation where this
%difference may be important.
%%This is the place to cite works suggesting collective enhancement
%\vspace{0.2cm}

\noindent
Let us consider how a parity violating energy difference could
manifest itself in the collision of two molecules in a BEC.
Remember that in order to form a chiral molecule there must be at
least four atoms involved \cite{four}; therefore the collision of two diatomic
molecules is sufficient.
The distance between energy levels in a combined molecule $D$
is much smaller than the distance in an initial molecule, $D \ll D_{i}$.
Therefore, in an ``ordinary'' system it would be impossible to
resolve compound resonances of the combined system since we would
expect $k_{B}T \gg D$, where $k_{B}$ is the Boltzmann constant,
$T$ is the temperature.
However, in a molecular BEC a finite fraction of molecules (up to 100\%
for $T \ll T_c$) have
zero kinetic energy, $E=0$ \cite{commentE=0}.
This means that we have a possibility of  perfect energy resolution
 \cite{internal}.
This allows one to study molecular collisions and chemical reactions
in a unique situation where the reaction is dominated by the
closest compound resonance of the combined system.
\vspace{0.2cm}

\noindent
We can express the cross-section for formation of a chiral compound
molecular state by the Breit-Wigner formula
\begin{equation}
\sigma = \frac{\pi}{k^{2}}\frac{\Gamma _{c}\Gamma}
{(E-E_{0})^{2}+\Gamma ^{2}/4} \ ,
\end{equation}
where $k$ is the wave vector, $\Gamma _{c}$
is the capture width, and $\Gamma$ is the total width of the
resonance.
The parity violating weak interaction in the chiral molecules shifts the
resonance energies, for example, let's consider that for the left-handed
structure $E_{0}\rightarrow E_{L}=E_{0}-\Delta E_{PV}/2$ while for the
right-handed structure
$E_{0}\rightarrow E_{R}=E_{0}+\Delta E_{PV}/2$.
Therefore, cross-sections for the formation of left
and right molecules, $\sigma _{L}$ and $\sigma _{R}$, from
achiral components may be different.
We can define an asymmetry parameter
\begin{equation}
P=\frac{\sigma_{L}-\sigma _{R}}{\sigma _{L}+\sigma _{R}} \ .
\end{equation}
The maximum value for $P$ is reached when $E=E_{0}\pm \Gamma/2$.
At this energy the asymmetry parameter
\begin{equation}
|P_{\rm max}|=\frac{\Delta E_{PV}}{\Gamma} \ .
\end{equation}
The resonances can be shifted close to zero energy
(the energy of the molecular collisions)
$E_{0}\pm \Gamma/2=0$ by application of an external
electric or magnetic field.
\vspace{0.2cm}

\noindent
The width of the level $\Gamma$ may be quite small since the capture
width $\Gamma_c=0$ for energy $E=0$.
The radiative width may be smaller than $10^7 {\rm Hz}$.
This can be compared to the weak energy shift. The largest
values for $\Delta E_{PV}$ that have been considered in molecular
calculations are $\sim 10^4 {\rm Hz}$
(e.g., for H$_{2}$Po$_{2}$ \cite{molcalcs}).
%remember that the energy difference reduces rapidly when
%lighter elements are substituted for Po).
\vspace{0.2cm}

%\noindent
%To test this idea one can first search for similar effects produced by
%the application of other chiral fields, for example,
%crossed magnetic and electric fields
%(effect ${\bf E}\times {\bf B}$).
%In this case the energy difference can be much larger than $\Delta E_{PV}$
%and can be controlled by varying the strength of the fields.
%\vspace{0.2cm}

\noindent
Another possibility to observe parity violation in molecules may be
related to the admixture of an $s$-wave to a $p$-wave compound resonance.
For energy $E=0$, only $s$-wave molecules have a significant
interaction cross section. Consider now a $p$-wave compound resonance.
It seems to be invisible for $E=0$.
However, the weak interaction $W$ mixes states of opposite parity
and produces the combined state $|\psi\rangle=|p\rangle +\beta|s\rangle$,
thus opening the $s$-wave reaction amplitude proportional to $\beta$.
The mixing coefficient  $\beta=\langle p | W |s\rangle/(E_s-E_p)$
 is enhanced since the energy interval between the opposite parity
compound states $(E_s-E_p)$
is very small due to the high level density in a combined molecule.
Note that this mechanism is responsible for the enhancement of
weak interaction effects in neutron-nucleus reactions $\sim 10^{6}$
times \cite{sf} (for a review of the experiments, see \cite{nNexp}).
%A similar effect can be achieved by an external electric field
%$\varepsilon$. The interference between the weak and electric amplitudes
%(the effect linear in $\varepsilon$) may be a signature of the weak
%interaction.
Interference of the very small $p$-wave amplitude
and the weak-induced $s$-wave amplitude may possibly also lead to a
difference in the production of right and left molecules in the $p$-wave
resonance.
\vspace{0.3cm}

\noindent
This work was supported by the Australian Research Council.

%*************************************************************************


\begin{thebibliography}{20}
\bibitem{comment2heavy}

In such molecules $\eta$ can be many orders of magnitude
larger than in molecules with one heavy atom \cite{RHS}.
\bibitem{four}

Note that it is not necessary to have all four atoms different.
Even such simple molecules as H$_2$O$_2$ have a chiral structure.
\bibitem{commentE=0}

Strictly speaking, the kinetic energy is not exactly zero
due to the finite size of the system.
The ground state of the molecules, however, is very close
to $E=0$ and the kinetic energy spread is vanishingly small
($\Delta E \sim \hbar ^2 /M R^2$, where $M$ is the molecular
mass, $R$ is the size of the system).
\bibitem{internal}

Colliding molecules can be in an internal excited state.
However, if $D_i > k_{B}T$ the probability of the excited state
is small. Moreover, the molecules which are not in the ground state
will probably not be trapped and will quickly leave the system.
\end{thebibliography}
\end{document}